\documentstyle[12pt,eqsecnum,multicol,epsfig,aps,prl,array]{revtex}

\draft
\widetext
\begin{document}
\title{Rigidity transitions and constraint counting in amorphous networks: 
beyond the mean-field approach} 
\author{Matthieu Micoulaut}
\address{Laboratoire de Physique Th{\'e}orique des Liquides,
Universit{\'e} Pierre et Marie Curie, Boite 121\\
4, Place Jussieu, 75252 Paris Cedex 05, France\\}

\date{\today}
\maketitle
\begin{abstract}
\par
We study rigidity transitions in covalent 
amorphous  networks using
size-increasing cluster approximations and constraint counting algorithms.
Possible consequences of the presence of self-organization are examined. 
The analysis reveals two
transitions instead of the usual (mean-field) single transition. One from 
a floppy 
to an isostatic rigid phase at a mean coordination number $\bar r_{c1}$
where the number of floppy modes vanishes and a second one from an
isostatic to a stressed rigid phase 
at $\bar r_{c2}$. The value of the
two critical mean coordination numbers as well as the width $\Delta \bar r
=\bar r_{c2}-\bar r_{c1}$ of the
intermediate phase depend very strongly on the presence of medium range
order elements such as rings.
\par
{Pacs:} 61.43.Fs-46.30.Cn
\end{abstract}
$$ $$
The notion of constraints and their application to classical
macroscopic physics problems such as the stability of
bridges and trusses have been introduced and first considered
by J.L. Lagrange and 
J.C. Maxwell \cite{Lagrange,Maxwell}.  
On this basis, J.C. Phillips asserted\cite{Phillips1,Phillips2} some
twenty years ago that
covalent networks can
be mechanically constrained by interatomic valence forces such as bond-
stretching and bond-bending and optimal glass formation is
attained when the netwok sits at a
mechanically critical point. This happens when the constraints $n_c$ per
atom  estimated by Maxwell counting equal
the degrees of freedom per atom in 3D, i.e. $n_c=3$. 
\par
Such mechanical systems have been examined in
terms of percolation theory by M.F. Thorpe\cite{Thorpe} who 
showed by a normal mode analysis that the number of zero 
frequency solutions (floppy modes) $f$ of the dynamical matrix 
equals $f=3-n_c$ and vanishes when the mean coordination number
$\bar r$ of the network reaches the critical value $\bar
r_c=2.4$. In this mean-field approach, one considers a network of N
atoms composed of $n_r$ atoms that are $r$-fold coordinated. The
enumeration of mechanical constraints in this system gives $r/2$
bond-stretching constraints and $(2r-3)$  bond-bending constraints
for a $r$-fold coordinated atom. Since then, a certain number of 
structural possibilities
have been taken into account such as rings, broken bond-bending 
constraints\cite{Science}
or the effects of one-fold coordinated atoms\cite{Boolthorpe}.  
These powerful ideas have led to the prediction of a
floppy to rigid transition in random 
networks and various examples where rigidity percolation threshold
occurs have been reported\cite{r8}. Also, applications of rigidity 
in biology and computational science have been reported\cite{biol},
\cite{comp}.
Nevertheless, experiments on binary 
and ternary chalcogenide glasses have  shown the 
existence
of two transitions at $\bar r_{c1}$ and $\bar r_{c2}$ instead of 
the single mean-field transition\cite{Bool1}-\cite{Bool3}. 
This suggests that the 
mean-field constraint counting alone as it has been realized up to
now, may 
be insufficient to describe accurately the underlying phase
transitions. Recent results on  
Raman scattering and modulated differential scanning calorimetry (MDSC)
realized on Group IV binary chalcogenides or
ternary glasses suggest indeed evidence\cite{Bool4}
for the growth of a 
self-organized (isostatic rigid) intermediate phase between 
the floppy and the stressed rigid phases, for which evidence
is obtained from numerical simulations\cite{Thorpe01}.   
\par

However, a certain number of questions remain at this stage. What controls 
the values $\bar r_{c1}$ and $\bar r_{c2}$, the width $\Delta \bar r=\bar
r_{c2}-\bar r_{c1}$ ?
Recent results show that this width can be particularly sharp\cite{PRL01}.
How does isostatic regions and self-organization influence the absolute 
magnitude of these quantities ? What can be done which goes beyond the
elegant mean-field approach ? This Letter attempts to adress these basic
issues. We report here on the role of medium range order (MRO) in
glasses of the form $B_xA_{1-x}$ with coordination numbers $r_A=2$ and
$r_B=4$, and $\bar r=2+2x$. Typical glasses are the Group IV
chalcogenides such as
$Ge_xSe_{1-x}$ which have been extensively studied in this context. To
construct 
MRO, we have used size-increasing cluster
approximations (SICA) to generate sets of clusters on which we have
applied constraint counting algorithms. The results show two
transitions, one at which the number of floppy modes vanishes.
Another transition (a ``stress transition'')  where
stress in the structure can not be avoided anymore, is located
beyond. In between, this
provides evidence for a self-organized network for which the
probability of stress-free clusters has been computed. The width $\Delta \bar
r$ increases with the fraction 
of MRO elements. Finally, in case of random bonding, a single transition is 
obtained.
\par
{\em Construction.} SICA have been first introduced to elucidate the
formation of fullerenes\cite{Kerner1}, but also the intermediate range
order in amorphous
semi-conductors\cite{Micoul}. They rely on the statement that the
fraction of significant MRO structures converges very rapidly to a limit value
when the size of the considered clusters is increasing\cite{limit}. 
The construction is
realized in Canonical Ensemble with particular energy levels. One starts from
short range order molecules (the basic units at the initial step
$l=1$ which will serve as building blocks) and construct all
possible structural arrangements of two basic units ($l=2$, see
Table I), three basic units ($l=3$) and so on. This is supposed to be
realized at the formation of the network, when T equals the fictive 
temperature $T_f$ \cite{Galeener}. Here, we have chosen as basic units
the $A_2$ and the
stoichiometric balanced $BA_2$ molecules (e.g. $Se_2$ and $GeSe_2$) for a
reason which will become clear below. We have checked that the results
do not depend on this particular initial choice. These basic units
have respective probabilities $1-p$ and $p=2x/(1-x)$, $x$
being the concentration of B atoms.
\par 

The creation of a
chain-like $A_2-A_2$ structure will involve an energy gain of $E_1$,
isostatic $A_2-BA_2$ bondings will use an energy gain of $E_2$ and the
creation of corner-sharing (CS) and edge-sharing (ES) $BA_{4/2}$
tetrahedra respectively $E_3$ and $E_4$. The latter quantity will be
used to tune the fraction of ES among the structure. The produced
probabilities have different statistical
weights which corresponds to the number of equivalent ways a given
cluster can be constructed. This quantity can be regarded as the
degeneracy of the corresponding energy level. For instance, given the
coordination number $4$ of the basic unit $BA_{4/2}$ and labeled
covalent bonds a CS $B_2A_4$ cluster has the multiplicity $4\times 4$,
whereas for a ES cluster, we count $2\times \left( \begin{array}{c}4\\2
\end{array}\right)  \left( \begin{array}{c}4\\2
\end{array}\right)=72$ in three dimensions

\par
Due to the initial choice of the basic units, the value of the energy
$E_2$ will influence the probability of isostatic clusters since this
quantity is involved in the probability of the isostatic $BA_4$
cluster
($n_c=3$, see Table I). If
we have $e_2\gg e_1,e_3,e_4$, the network will be mainly isostatic.
\par
At step $l=2$, we can generate three types of clusters (Table I),
$A_4$, $BA_4$ and $B_2A_4$ having two isomers (the CS and ES
tetrahedra). Their unrenormalized probabilities are given by:
$p_{A_4}=4(1-p)^2e_1$, $p_{BA_4}=16p(1-p)e_2$, $p_{CS}=16p^2e_3$ and $p_{ES}=72p^2e_4$
out of which can be extracted the concentration $x^{(l=2)}$ of $B$ atoms.
The quantities $e_i=exp[-E_i/T_f]$ are the Gibbs weights at
$T_f$. Next, we compute the number of mechanical constraints
(bond-bending and bond-stretching) per atom
on each cluster by Maxwell counting. Special care has to be taken in 
order to avoid the counting of redundant constraints on clusters
containing rings, following the procedure described by Thorpe\cite{Thorpe}.
The probabilites depend on two parameters (i.e. the Gibbs weights
$e_1/e_2$ and $e_3/e_2$) and eventually $e_4/e_2$ if one considers
the possibility of ES or rings. One of
these two weights can be calculated by writing a
conservation law for the
concentration of $B$ atoms\cite{Bray}:
\begin{equation}
\label{1}
x^{(l)}=x
\end{equation}
These weights become composition dependent in solving equ. (\ref{1})
which means that either the energies $E_i$ or the fictive temperature
$T_f$ depend on $x$\cite{Galeener} but here only the $e_i(x)$
dependence is relevant for our purpose.
With increasing cluster size, it is obvious that the number of
potential isomers will increase (Table I), also the different types of
rings which
have some evidence in chalcogenides\cite{GeSe}.
We have realized the construction up to the step $l=4$.
At each step,
we have determined either $e_1/e_2$ or $e_3/e_2$ solving
equ. (\ref{1}) and computed the total number
of constraints $n_c$ per atom on the set of clusters (see Table I). Finally, we 
have looked for the concentration of $B$ atoms (or the mean coordination
number $\bar r$) for which the number of floppy
modes vanishes.
\par
{\em Results.} Random bonding is obtained by setting the above defined
Gibbs weights
$e_i$ to one and the cluster probabilities are then only given by
their statistical weights.

\par
Solving  $f=0$, one obtains a single  transition for all steps in the
mean coordination number  range [2.231,2.275], somewhat lower than
the usual mean-field value 2.385\cite{Thorpe01}.
This comes from  the fact that the number
of equivalent ways  to connect $BA_{4/2}$  units     together is
substantially  higher than   for  the connection  of  (chain-)  $A_2$
units. We  do not obtain an  intermediate  phase for random bonding.
\par
Let us turn to self-organization and proceed as follows. Starting from a floppy
cluster of size $l$ (almost a chain-like structure made of A atoms), 
we allow the agglomeration of a new
basic unit onto this cluster to generate the cluster of size $l+1$ 
only if the creation of a stressed rigid region can be avoided
on this new cluster (due to the agglomeration of a $BA_{4/2}$ basic unit
onto another $BA_{4/2}$ tetrahedron being part of the $l$-sized
cluster). With this rather simple rule, upon increasing $\bar r$ we will
accumulate isostatic rigid regions on the size increasing clusters
because $BA_{4/2}$ units are only accepted in 
$A_2-BA_{4/2}$ isostatic bondings. Alternatively, we can start from a
rigid cluster
which exist at higher mean coordination number ($\bar r\leq 2.67$) and
follow the same procedure but in opposite way, i.e. we allow only
bondings which lead to isostatic rigid regions, excluding
systematically the possibility of floppy $A_2-A_2$ bondings.
\par
Here, the simplest case deals with  dendritic clusters, where we have 
removed all
possibilities of ring creation. For $l\to \infty$, this would permit to
recover the results on 
Random Bond Models\cite{RBM} for which there are no loops or rings in the
thermodynamic limit 
and to obtain equivalence with Bethe lattice solutions\cite{Bethe}. We
obtain a single transition  for even $l$ steps at exactly the
mean-field value $\bar r=2.4$ whereas for the step $l=3$, there is
a sharp intermediate phase defined by $f=0$ (again at $\bar r=2.4$) and the 
vanishing of floppy regions (i.e. $e_1/e_2$ is zero) at $\bar
r=2.382(6)$. The probability of isostatic clusters as a function of the mean
coordination number has been computed and shows
that the network is entirely stress free at the point where $f=0$
(solid line, fig. 2). If
there is a width (for $l=3$), then the same probability is less than one and
displays a narrow distribution.
\par
Next, we have allowed a certain amount of medium range order (MRO) by
setting the quantity $e_4/e_2\neq 0$. Two transitions are then obtained for
every SICA step. A
first one at $\bar r_{c1}$ where the number of floppy modes
vanishes. A second one
at $\bar r_{c2}$ defined by $e_3/e_2=0$. This means that
beyond this point, stressed rigid regions created by the connection of at least
two $BA_{4/2}$ units can not be avoided anymore, i.e. the Gibbs weight
$e_3/e_2$ becomes non-zero, and composition dependent. We call this
point the ``stress
transition'' because its definition is very close to the one given in
\cite{Thorpe01}. We show
the $l=2$ result (fig. 2) where $f=0$ at $\bar r_{c1}=2.4$ and $e_3\neq
0$ at $\bar r>\bar r_{c2}$ for different fractions of ES tetrahedra, defining the
intermediate phase $\Delta \bar r$. $\bar r_{c1}$
does not depend on the ES fraction, as well as the fraction of
stressed rigid clusters
in the structure. To ensure continuous deformation of the network when
B atoms are added and keeping the sum of the probability of floppy, isostatic
rigid and stressed rigid clusters equal to one, the probability of isostatic 
rigid clusters connects the isostatic solid line at $\bar r_{c2}$. Stressed rigid
rings first appear in the region $\bar r_{c1}<\bar r<\bar r_{c2}$
while chain-like stressed clusters (whose probability is proportional
to $e_3$) occur only beyond the stress
transition, when $e_3\neq0$. We conclude that when $\bar r$ is increased, stressed
rigidity nucleates through the network starting from rings. Results
remain similar for the even $l=4$ step. It appears from fig. 2 that
the width $\Delta\bar r=\bar r_{c2}-\bar r_{c1}$ of
the intermediate phase increases with the fraction of MRO. We have
represented this quantity as a function of the MRO fraction at the
rigidity transition in fig. 3 
which shows that $\Delta \bar r$ is almost an increasing function of the ES
fraction as seen from the result at SICA step $l=4$. Here, there is only a
small difference between allowing only four-membered rings (ES) (lower
dotted line) or
rings of all sizes (upper dotted line) in the clusters. Finally, one
can see from
fig. 2 and the insert of fig. 3 that the probability of isostatic clusters is
maximum in the window $\Delta \bar r$, and almost equal to 1 for
the even SICA steps, providing evidence that the structure is almost 
stress-free.  
\par
{\em Discussion.} Chalcogenide glasses represent the ideal systems to check
these results. Different types of experimental measurements have
given evidence on the two transitions and the nature of the
self-organized intermediate phase. Raman scattering has been used
\cite{Bool1,Bool3} as a
probe to detect elastic thresholds in $Si_xSe_{1-x}$ and $Ge_xSe_{1-x}$
glasses. Specifically, changes in the CS mode chain frequencies have
been studied with glass compositions and show a kink (or a jump) at
the mean coordination number $\bar r_{c1}=2.4$ and $\bar
r_{c2}=2.52$ in Ge and $\bar r_{c2}=2.54$ in Si based systems,
suggesting onset of a new rigidity at $\bar r_{c2}$.
A clear correlation between these results and the vanishing of the
non-reversing heat flow $\Delta H_{nr}$ (the part of the heat flow
which is thermal history sensitive) in MDSC measurements has been
shown\cite{Bool1,Bool3}. Obviously, since this $\Delta H_{nr}$ term
provides a measure of how
different a glass is from a liquid in a configurational sense, this
suggest that in the intermediate phase, glass and liquid structure are
closely similar to each other.
\par
The SICA and constraint counting algorithms show that the width $\Delta\bar
r$ of the intermediate phase increase with the fraction of ES
tetrahedra and more generally with MRO composed of small rings
(fig. 3). We stress that the width should converge to a lower limit
value of $\Delta \bar r$ compared to the step $l=2$, therefore one can
observe the shift downwards when increasing $l$ from 2 to 4. This
limit value is in principle attained for $l\to \infty$, or at least for much
larger steps than $l=4$ \cite{Micoul}. For Si-Se, $\Delta\bar r=0.14$ is
somewhat larger than forGe-Se ($\Delta\bar r=0.12$) consistently with the
fact that the number of
ES is higher in the former\cite{Bool3}.  
\par
In summary, we have shown that size increasing cluster approximations
could be used to go beyond the mean-field approach of the rigidity
transitions. We have estimated for the different approximation steps
the number of mechanical constraints and the number of floppy modes $f$
and shown that two transitions were obtained in this situation. One at
which $f$ vanishes and another at which stressed rigid regions appear
on the clusters. The width $\Delta \bar r$ is an increasing function of
the MRO fraction. In the window $\Delta \bar r$,
the rate of isostatic clusters is at its maximum. These new results should
motivate developments on the role of local structure and MRO in
the rigidity transition, and applications to Group
V chalcogenides such as $As_xSe_{1-x}$ glasses.
\par
The Laboratoire de Physique Th{\'e}orique des Liquides is Unit{\'e} Mixte de
Recherche n. 7600 du CNRS.

\newpage
\listoffigures
\listoftables
\newpage
\begin{figure}
\begin{center}
\epsfig{figure=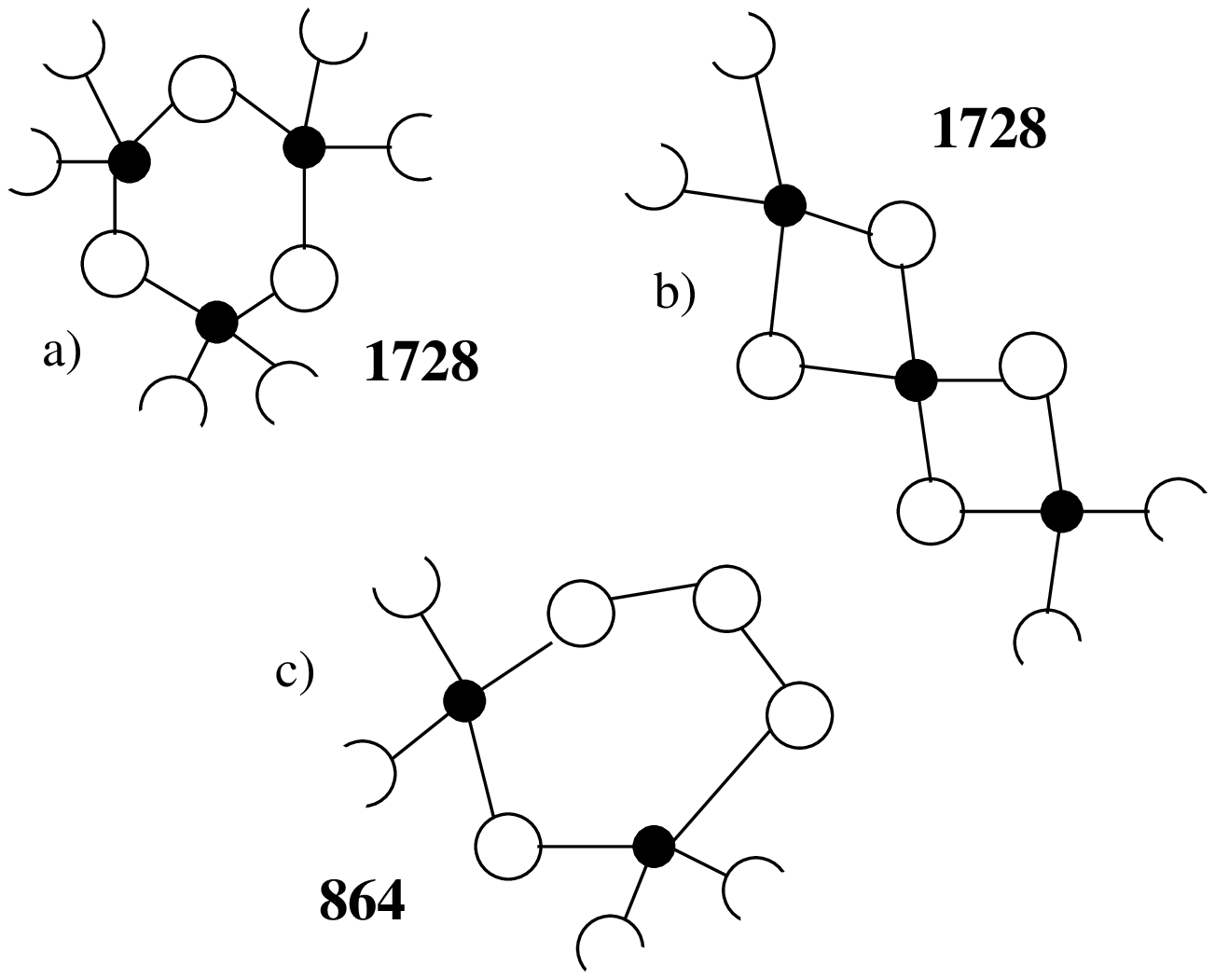,width=10cm}
\caption{Some of the $Ge_3Se_6$ and $Ge_2Se_6$ MRO clusters generated
by SICA at step $l=3$ with
their statistical weight. a) A six-membered ring with $n_c=3.67$ b) An
edge-sharing tetrahedra chain typical of vitreous $SiSe_2$ with
$n_c=3.22$ c) A six-membered ring with chalcogen inclusions and $n_c=3.25$.}
\end{center}
\end{figure}
\newpage
\vspace{0.8cm}
\begin{figure}
\begin{center}
\epsfig{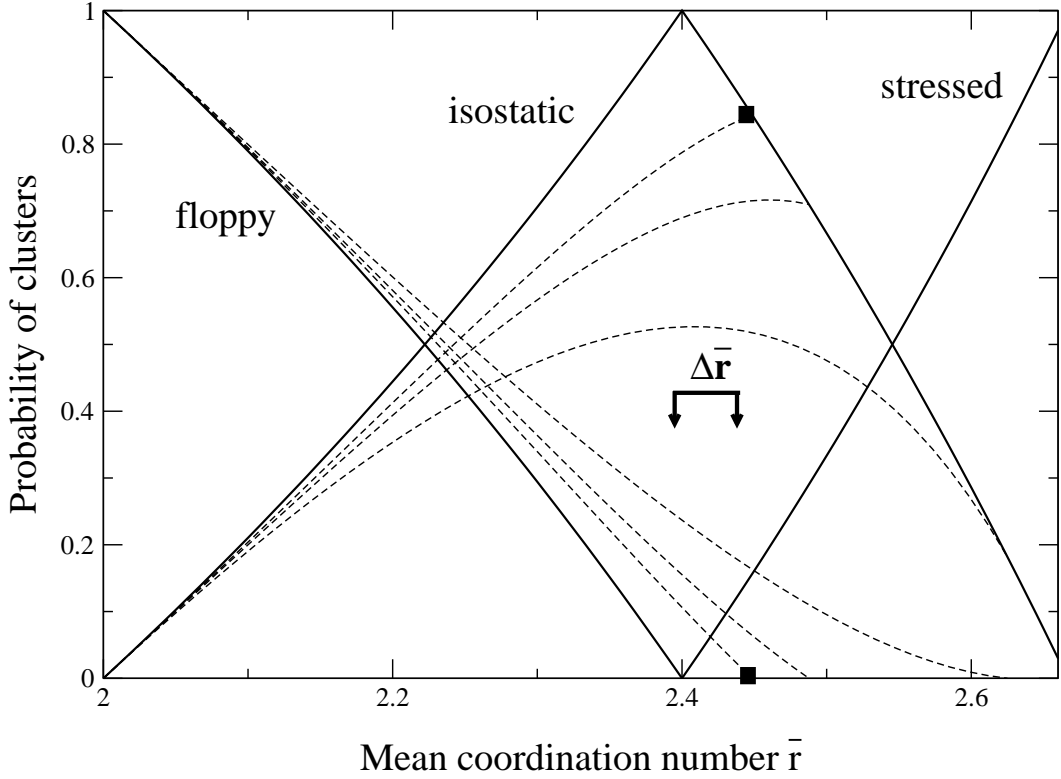}
\vspace{0.5cm}
\caption{Probability of floppy, isostatic rigid and stressed rigid
clusters as a
function of the mean coordination number for different fractions of
ES at $l=2$. The solid lines correspond to the dendritic
case where no edge-sharing tetrahedra are allowed. The broken lines
correspond to the same quantities for ES fraction at the stress transition
of 0.156, 0.290 and 0.818. For a ES fraction of 0.156, the filled squares 
indicate the point $\bar r_{c2}$ at which the stress transition occurs
and serves to define the intermediate phase $\Delta
\bar r$.}
\end{center}
\end{figure}
\newpage
\vspace{0.5cm}
\begin{figure}
\begin{center}
\epsfig{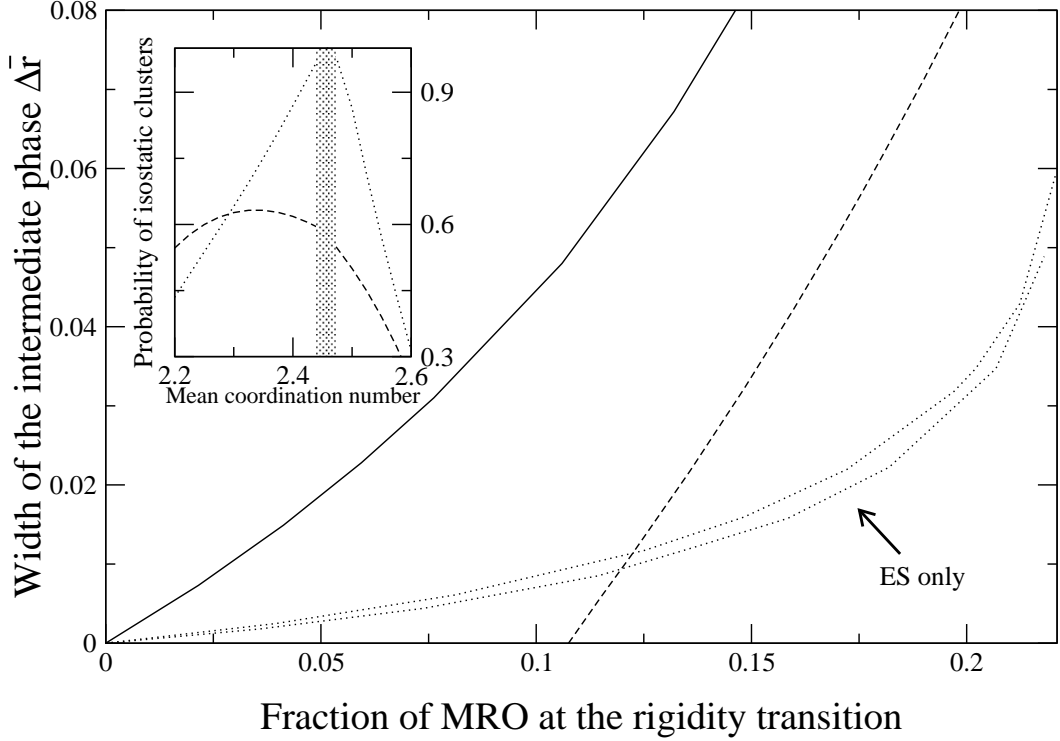}
\vspace{0.5cm}
\caption{Width of the transition $\Delta \bar r$ as a function of the fraction
of MRO clusters at the rigidity transition for $l=2$ (solid line),
$l=3$ (dashed line) and $l=4$ (dotted lines). At step $l=2$, the MRO clusters reduce to the edge-sharing 
$GeSe_{4/2}$ tetrahedra. For larger steps, different rings sizes
($4,6,8$) have
been taken into account. The lower dotted line correspond to a system
at $l=4$
having only ES as MRO element. The insert shows the probability of isostatic
clusters with mean coordination number $\bar r$ for $l=4$ (dotted
line) and $l=3$ (dashed line). The shaded region of $l=4$ is defined
by the corresponding $\Delta \bar r$.}
\end{center}
\end{figure}
\newpage
\vspace{0.5cm}
\begin{table}
\begin{center}
\begin{tabular}{ccccc}\hline\hline
Size $l$&cluster&Number of isomers&$n_c$ \\ \hline
1&$Se_2$&1&2 \\
 &$GeSe_2$&1&3.67 \\ \hline
2&$Se_4$&1&2 \\
   &$GeSe_4$&1&3 \\
   &$Ge_2Se_4$&2(1)&3.67 \\ \hline
3&$Se_6$&1&2 \\
   &$GeSe_6$&2&2.71 \\
   &$Ge_2Se_6$&4(2)&3.25 \\
   &$Ge_3Se_6$&4(3)&3.67 \\ \hline
4&$Se_8$&1&2 \\
   &$GeSe_8$&3&2.56 \\
   &$Ge_2Se_8$&11(6)&3 \\
   &$Ge_3Se_8$&12(9)&3.36 \\
   &$Ge_4Se_8$&10(9)&3.67 \\ \hline\hline
\end{tabular}
\end{center}
\caption{Clusters generated at the different SICA steps $l$ with the
chemical formula in case of $Ge_xSe_{1-x}$ glasses, the number of
isomers and the number of constraints $n_c$ per atom. The number of
clusters containing rings is  indicated in bracketts. $GeSe_4$ and
$Ge_2Se_8$ are isostatic clusters with respective energy levels $E_2$
and $2E_2$.}
\end{table}

\begin{thebibliography}{100}
\bibitem{limit} In fullerenes, one has 12 pentagons and 32
hexagons. The statistics of
pentagons and hexagons in clusters of size $n<60$ must therefore converge
to the corresponding obvious limit values. In vitreous $B_2O_3$, the
fraction of three-membered rings should converge to 0.83, according to
experiment.
\end{thebibliography}
\end{document}